\documentclass[article,reprint,amsmath,amssymb,superscriptaddress]{revtex4-1}
\usepackage[colorlinks,
linkcolor=blue,
anchorcolor=blue,
citecolor=blue,
]{hyperref}
\usepackage{graphicx}% Include figure files untitled
\usepackage{epstopdf}
\usepackage{hyperref}
\UseRawInputEncoding
\setcitestyle{super} %superscript
\usepackage{breakurl}
\usepackage{multirow}
\usepackage{enumitem}
\usepackage{color}

\usepackage{tablefootnote}
\usepackage{threeparttable}
\usepackage{tabularx}
\usepackage{booktabs}
\usepackage{textcomp}
\usepackage{lineno}
\begin{document}

\title {AV$_3$Sb$_5$ kagome superconductors: a review with transport measurements}

\author{Zhuokai Xu}
\affiliation{Key Laboratory for Quantum Materials of Zhejiang Province, Department of Physics, School of Science and Research Center for Industries of the Future, Westlake University, Hangzhou 310030, P. R. China}
\affiliation{Institute of Natural Sciences, Westlake Institute for Advanced Study, Hangzhou 310024, P. R. China}

\author{Tian Le}
\affiliation{Key Laboratory for Quantum Materials of Zhejiang Province, Department of Physics, School of Science and Research Center for Industries of the Future, Westlake University, Hangzhou 310030, P. R. China}
\affiliation{Institute of Natural Sciences, Westlake Institute for Advanced Study, Hangzhou 310024, P. R. China}

\author{Xiao Lin}
\email{linxiao@westlake.edu.cn}
\affiliation{Key Laboratory for Quantum Materials of Zhejiang Province, Department of Physics, School of Science and Research Center for Industries of the Future, Westlake University, Hangzhou 310030, P. R. China}
\affiliation{Institute of Natural Sciences, Westlake Institute for Advanced Study, Hangzhou 310024, P. R. China}

\date{\today}

\begin{abstract}
\noindent
Kagome systems have garnered considerable attention due to the unique features of the sublattice structure and band topology.  
The recently discovered kagome metals AV$_3$Sb$_5$ (where A = K, Rb, Cs) host a rich array of symmetry-breaking phases, including exotic charge density waves (CDW), electronic nematicity, pair density waves (PDW) and superconductivity. Despite extensive experimental and theoretical investigations into the diverse phases,  several key issues remain contentious, such as a solid clarification of the time-reversal symmetry breaking (TRS-breaking) in the CDW order and its implications for the nature of superconducting (SC) pairing symmetry. This review aims to shed light on the transport properties of these intertwined phases, emphasizing the pivotal role that transport measurements play in uncovering the non-trivial quantum states of matter.

\end{abstract}

\maketitle

\begin{figure*}[thb]
	\begin{center}
		\includegraphics[width=18cm]{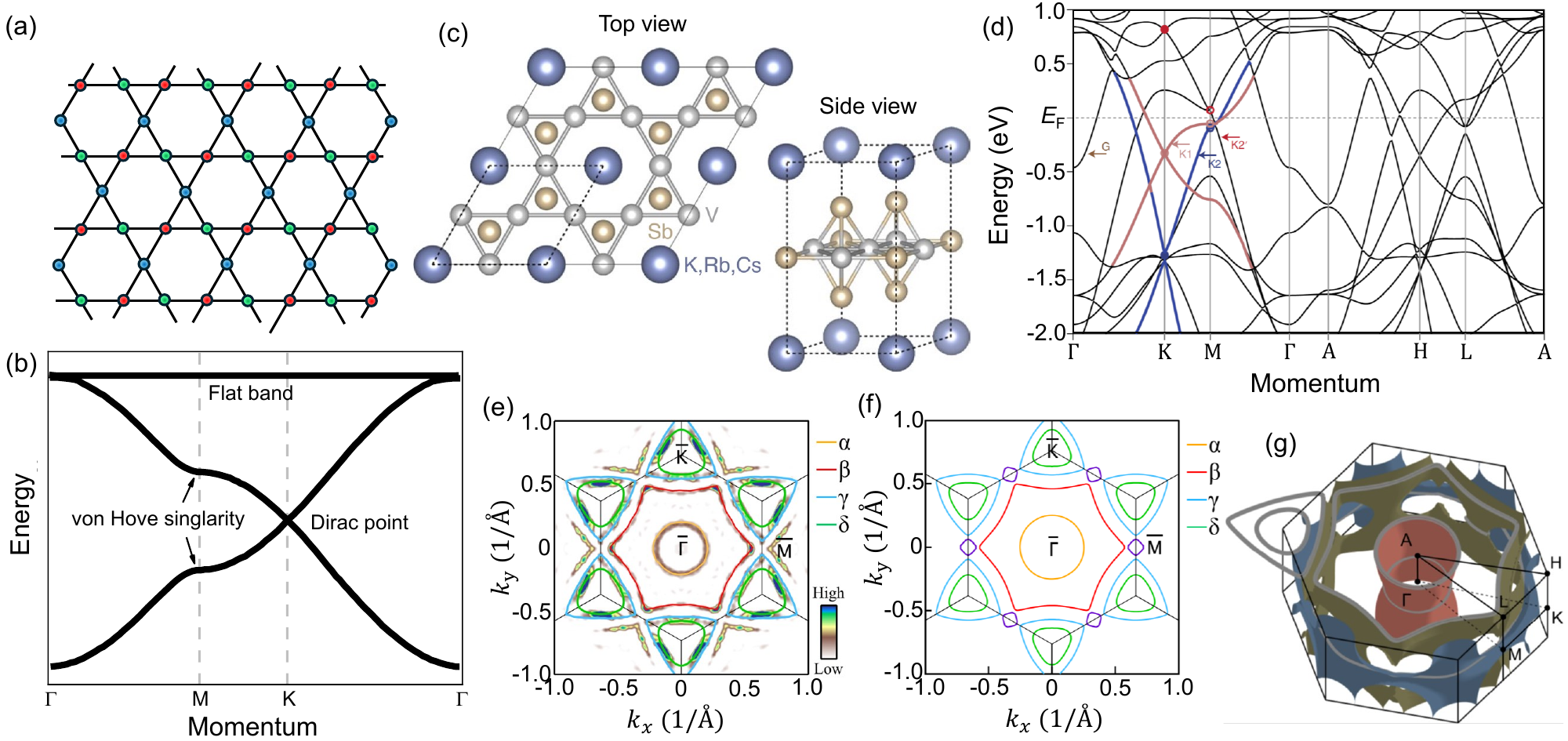}
	\end{center}
	\setlength{\abovecaptionskip}{-8 pt}
	\caption{(a) Kagome Lattice. The three sublattices are marked by different colors. (b) Band structure of kagome lattice from the nearest-neighbor tight-binding model. (c) Crystal structure of AV$_3$Sb$_5$~\cite{Comin2023NM}. (d) Calculated band structure of CsV$_3$Sb$_5$~\cite{Comin2023NM}, showing four bands ($G$, $K1$, $K2$ and $K2'$) crossing Fermi level. vHSs at $M$ and Dirac points at $K$ stems from $K1$, $K2$ and $K2'$ kagome bands and are marked by open and solid circles, respectively.		
	(e) Fermi surface mapping of KV$_3$Sb$_5$ from ARPES measurements~\cite{ZhouXJ2022NC}. (f) Calculated Fermi surface at $k_\mathrm{z}=\pi/c$ for undistorted KV$_3$Sb$_5$~\cite{ZhouXJ2022NC}. (g) 3D Fermi surface for undistorted CsV$_3$Sb$_5$~\cite{Wilson2021PRX}. Here, (c) and (d) are taken from Ref.~\citenum{Comin2023NM}; (e) and (f) are taken from Ref.~\citenum{ZhouXJ2022NC}; (g) is taken from Ref.~\citenum{Wilson2021PRX}.
	}
	\label{Fig1}
\end{figure*}

\noindent \textit{1. Introduction.}
Introduced by Syozi in 1951~\cite{Syozi1951}, the kagome latttice is composed of corner-sharing triangles that form a hexagonal structure.  Each triangle contains three sublattices per unit cell, marked by different colors in Fig.~\ref{Fig1}a. This structure is geometrically frustrated. Given the nearest-neighbor antiferromagnetic exchange interaction, the kagome lattice, with its geometric spin frustration, serves an ideal platform for exploring quantum spin liquid~\cite{Balents2010Nature}. A standard tight-binding model on the kagome lattice yields a nontrivial electronic band structure, featuring two van Hove singularities at $M$-point for electron filling of $f = 5/12$ and $3/12$, a Dirac point at $K$-point for $f=1/3$ and a flat band~\cite{LiJX2012PRB,WangQH2013PRB,Thomale2013PRL}, as seen in Fig~\ref{Fig1}b. 
Consequently, by adjusting the electron filling, the kagome lattice exhibit a variety of emergent phases of matter, shaped by the combined effect of topology and strong correlations~\cite{Comin2020NC}. 
Once the filling approaches the van Hove singularity, the kagome model, incorporating on-site repulsion ($U$) and nearest-neighbor interaction ($V_1$), produces an exceptionally rich phase diagram, including charge bond-density wave~\cite{WangQH2013PRB,Thomale2013PRL}, spin density wave~\cite{LiJX2012PRB,WangQH2013PRB}, pair-density wave~\cite{Raghu2023PRB}, Z2 Chern insulator~\cite{Franz2009PRB} as well as unconventional superconductivity~\cite{LiJX2012PRB,Thomale2013PRL,WangQH2013PRB,Thomale2021PRL}.

%,ZengCG2022PRL,Checkelsky2024NP
%,Norman2016RMP,Zhou2017RMP,LinYP2021PRB,FengSP2022PRB,Thomale2012PRB,Andersen2022PRB
% i.e. the chemical potential,

%Depending on the band filling fraction n, these electronic states may engender various topological and correlated phases, as extensively investigated for more than a decade Commin NM 21–26

In 2019, Ortiz et al. reported the discovery of vanadium-based kagome metals, AV$_3$Sb$_5$ (A = K, Rb, Cs)~\cite{Ortiz2019PRM}. These compounds are layered materials, crystallizing in the P6/mmm space group. As illustrated in Fig.~\ref{Fig1}c. V-atoms constitute the kagome skeleton, with the V-Sb block responsible for the distinctive physical properties, while the alkali metal atoms serve as spacers between the layers. Later on, superconductivity was identified with the critical temperature, $T_\textrm{c} \approx 3$~K for CsV$_3$Sb$_5$~\cite{Wilson2020PRL} and $0.9$~K for KV$_3$Sb$_5$~\cite{Wilson2021PRM} and RbV$_3$Sb$_5$~\cite{LeiHC2021CPL}. Before entering the SC phase as temperature ($T$) lowers, AV$_3$Sb$_5$ undergoes a first-order CDW phase transition~\cite{Wilson2020PRL,Zeljkovic2021Nature,ChenXH2021PRX,WenHH2022PRB,Comin2023NM}, which is highly nontrivial~\cite{Guguchia2022Nature,Moll2022Nature,WuL2022NP,Hasan2021NM,Hasan2021PRB} and may represent the long-sought chiral flux phase~\cite{HuJP2021SB,Balents2022PRB}. Additionally, a reduction in rotational symmetry from C6 to C2 was observed within the CDW phase~\cite{Zeljkovic2022NP}, likely signaling the presence of electronic nematicity~\cite{ChenXH2022Nature,Zeljkovic2023NP}.  As $T$ further lowers below $T_\mathrm{c}$, a PDW phase emerges~\cite{GaoHJ2021Nature,YinJX2024Nature}, where the SC pairing amplitude modulates periodically in real space. These complex symmetry-breaking phases, either competing or intertwining with each other, is reminiscent of that observed in the high-$T_\mathrm{c}$ cuprates.

%Fernandes2022PRB,Zeljkovic2023PRX
% characterized by the doubling of lattice in the ab-plane , which breaks time-reversal symmetry (TRSB).

%ChenXH2022SCPMA,Luetkens2022PRR,WangZW2021PRB ~\cite{Zaanen2015Nature}. 

In the past few years, transport measurements have become an essential tool for probing the complex quantum phases of AV$_3$Sb$_5$. This review aims to emphasize their pivotal role and is organized as follows: we begin with an introduction to the basic electronic properties of AV$_3$Sb$_5$, drawing insights from quantum oscillation (QO) measurements. Next, we delve into the transport signatures of the CDW phase and the associated emergent symmetry breaking. Third, we present the current understanding of the SC order, highlighting nontrivial features such as time-reversal symmetry breaking (TRS-breaking), as unveiled by quantum transport techniques. Finally, we conclude with an outlook, addressing unresolved questions and suggesting potential future directions in the study of Kagome superconductors. This review may not cover all aspects of Kagome systems; for a broader understanding, one may refer to other insightful reviews~\cite{Hasan2022Nature,HuJP2023NSR,Ali2023NRP,HeMQ2023Tungsten,Wilson2024NRM,Yuan2024SST}.

%-----------------------------------------------------------
%------------------------------------------------------------

%(c) is reproduced with permission [36], Copyright 2023, Springer Nature Limited. (d) is reproduced with permission [37],
%Copyright 2021, The Authors. (e) is reproduced with permission [45], Copyright 2022, American Physical Society.
%where transport measurements may offer further insights.  covering angle-dependent resistance, elastoresistance, and field-switchable nonlinear resistance

%, which breaks time-reversal symmetry (TRSB).  of transport measurements and explore how they reveal the nontrivial properties that underpin the exotic phases of AV$_3$Sb$_5$. 

%The remarkable tunability of the 3D-CO state across otherwise similar compounds suggests that the AV 3 Sb 5 series is a candidate host for an extremely rich phase diagram of emergent electronic phases, enabling new opportunities for fundamental studies at the nexus of strong correlation phenomena and topology.

~\\
\noindent \textit{2. Basic electronic properties.}
The electronic band structure of AV$_3$Sb$_5$ retains key features of the ideal kagome model, but exhibits greater complexity~\cite{Wilson2020PRL,Wilson2021PRM,Miao2021PRX,Wilson2021PRX,ShiM2022SB,ShenDW2021PRL,ZhouXJ2022NC,Sato2021PRB}. Density functional theory (DFT) calculations reveal multiple bands near the Fermi level ($E_\mathrm{F}$)~\cite{Wilson2020PRL,Wilson2021PRX}, as shown in {Fig.~\ref{Fig1}d} for CsV$_3$Sb$_5$. An electron-like parabolic band emerges around the $\Gamma$ point, stemming from the Sb $p_z$-orbital. At the Brillouin zone (BZ) boundaries, the band structure is dominated by V $3d$-orbitals. Dirac points are at the zone corner $K$. And several van Hove singularities (vHSs) are observed at the the zone edge $M$, with two located close to $E_\mathrm{F}$. The Fermi surface nesting at these vHSs, combined with the divergence in the density of states (DOS), is believed to play a pivotal role in enabling pairing across multiple channels~\cite{LiJX2012PRB,WangQH2013PRB,Thomale2013PRL}. AV$_3$Sb$_5$ also exhibits a non-trivial $Z_2$ topological index~\cite{Wilson2020PRL}. Angle-resolved photoemission spectroscopy (ARPES) have identified the corresponding topological surface state, embedded around the bulk Fermi surface at $M$ point~\cite{ShiM2022SB}. 

{Fig.~\ref{Fig1}e-g} present a comparative study of experimental and calculated band dispersions and Fermi surfaces for AV$_3$Sb$_5$~\cite{Wilson2021PRX,Comin2023NM}. The Fermi pocket demonstrates a quasi-2D nature, indicating weak interlayer interactions between neighboring V-Sb blocks~\cite{Wilson2021PRX}.  The close agreement between ARPES and DFT calculations suggests only minor band renormalization, underscoring the weak correlation effects (small $U$) in the material~\cite{Yang2021PRB}.
Additionally, high resolution ARPES measurements have detected Fermi surface distortion induced by the CDW phase~\cite{ShenDW2021PRL,ZhouXJ2022NC,Sato2021PRB}. Kang $et~al.$ observed an unusual energy splitting in the vHS and Dirac bands, which is attributed to band folding in reconstructed BZ and points to a specific nature of the CDW phase (see more discussion below)~\cite{Comin2023NM}.

%This is reflected in the large resistivity anisotropy, with the ratio of out-of-plane to in-plane resistivity, $\rho_\mathrm{c}/\rho_\mathrm{ab}$, being around 600~\cite{Wilson2020PRL}.

%Borisenko2022PRL,WangSC2021PRX,Comin2022NP

%-----------------------------------------------------------
\begin{figure*}[thb]
	\begin{center}
		\includegraphics[width=18cm]{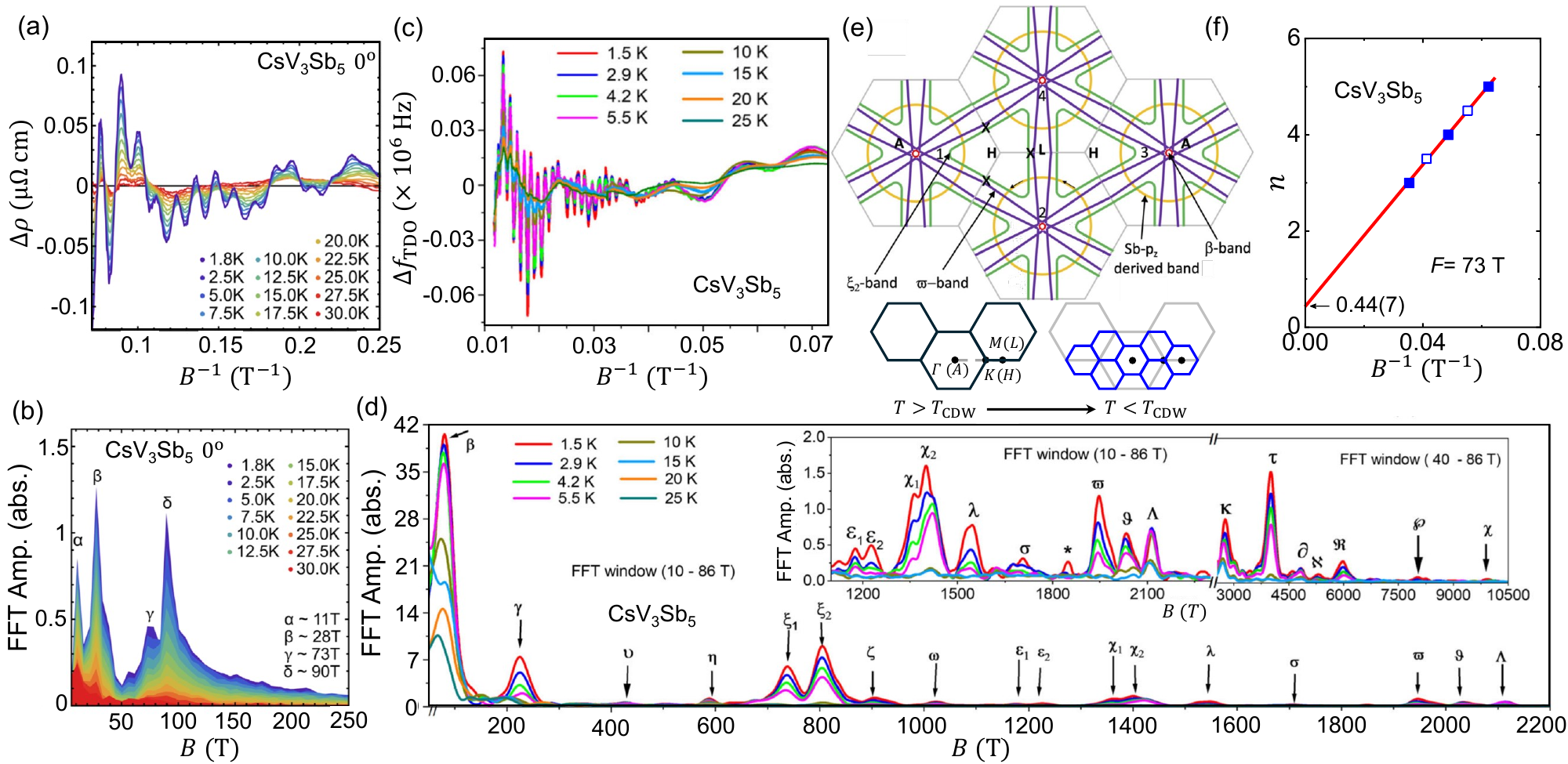}
	\end{center}
	\setlength{\abovecaptionskip}{-8 pt}
	\caption{(a) QOs of magnetoresistance measured up to 14~T for CsV$_3$Sb$_5$~\cite{Wilson2021PRX}. (b) Fourier transformation of the QO data in a, showing four frequency peaks below 250~T. (c) QOs of TDO frequency measured up to 86~T for CsV$_3$Sb$_5$~\cite{Welp2023PRL}. (d) Fourier transformation of the QO data in (c), showing multiple frequencies upto 10~kT. (e) Upper: Schematic of the 2$\times$2 folded Fermi surface of CsV$_3$Sb$_5$ $k_\mathrm{z}=\pi/c$ owing to CDW distortion~\cite{Welp2023PRL}. Lower: Evolution of BZs across the CDW distortion. (f) Landau fan diagram for the frequency of 73~T, revealing a nontrivial band topology~\cite{LeiHC2021PRL}. Here, (a-b) are taken from Ref.~\citenum{Wilson2021PRX}; (c-e) are taken from Ref.~\citenum{Welp2023PRL}; (f) are taken from Ref.~\citenum{LeiHC2021PRL}.
	}
	\label{Fig2}
\end{figure*}
%-----------------------------------------------------------

The multiple Fermi pockets in AV$_3$Sb$_5$ have intensively been investigated using QO measurements~\cite{Wilson2021PRX,LeiHC2021PRL,ChenXH2021PRB,Kwok2022PRB,HeMQ2023Tungsten,LiL2024NC,Welp2023PRL}.  When subjected to strong magnetic field ($B$), the electronic bands are quantized into discrete Landau levels. As $B$ increases, the Landau levels are progressively squeezed out of Fermi level, resulting in periodical oscillation in physical quantities associated with Fermi surface.  The QO period/frequency is indicative of the size of Fermi surface, specifically representing the extreme cross-section of Fermi pockets normal to $B$ (Onsager relation). Additionally, the $T$-dependence of QO amplitude  can be utilized to extract the effective electron mass ($m^*$) through the Lifshitz-Kosevich theory.
QO measurements are particularly powerful for detecting small Fermi pockets owing to the high energy resolution (sub-meV)~\cite{Lin2014PRL}, making it a crucial complement in the study of solid-state band structures.
The primary techniques of QOs include the quantum Hall effect, Shubnikov-de Haas (SdH) effect (magnetoresistance)~\cite{Wilson2021PRX,LeiHC2021PRL,ChenXH2021PRB}, and de Haas-van Alphen (dHvA) effect (magnetization)~\cite{Kwok2022PRB}. Besides that, QOs can be observed in phenomena such as the Seebeck effect~\cite{HeMQ2023Tungsten}, Nernst effect~\cite{HeMQ2023Tungsten}, thermal Hall effect~\cite{LiL2024NC} and tunnel diode oscillator (TDO)~\cite{Welp2023PRL}.

%,TianML2023Arxiv,HeMQ2022NJP,Goh2022PRB,RanS2022PRL
% beyond that of ARPES
%: $A(T)=\sinh(XT)/XT$, wherein $X=\frac{2\pi^2 k_\mathrm{B}}{\hbar e} \frac{m^*}{B}$ with $k_\mathrm{B}$ being the Boltzmann constant, $\hbar$ the reduced Plank constant and $e$ the charge unit. ($A$) 

\begin{figure*}[thb]
	\begin{center}
		\includegraphics[width=16cm]{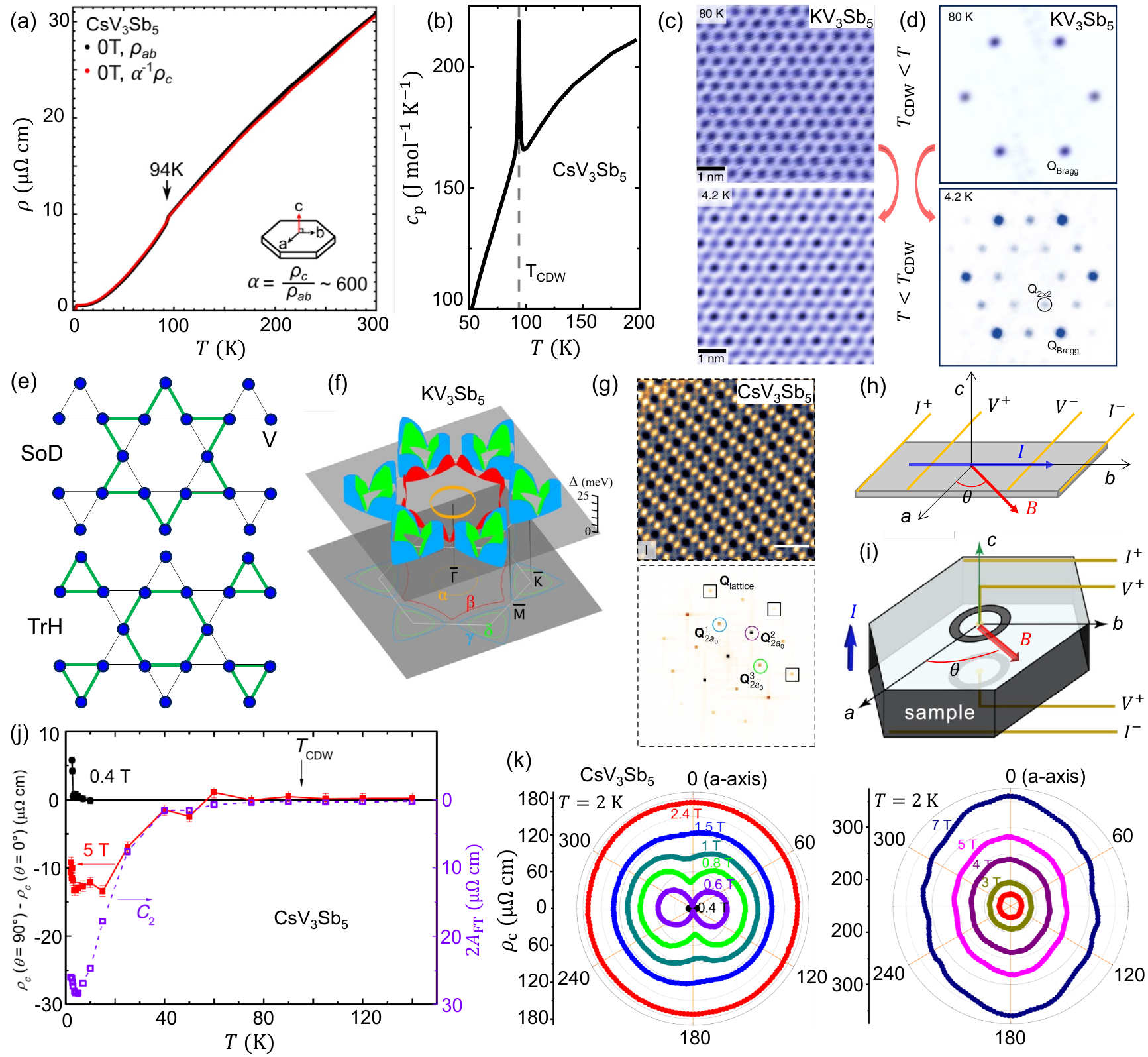}
	\end{center}
	\setlength{\abovecaptionskip}{-8 pt}
	\caption{(a) Temperature dependence of resistivity within in-plane and along c-axis for CsV$_3$Sb$_5$~\cite{Wilson2020PRL}. The anomaly at 94~K corresponds to the CDW transition. (b)  Temperature dependence of heat capacity~\cite{Wilson2020PRL}, showing a sharp peak at $T_\mathrm{CDW}\approx94$~K for CsV$_3$Sb$_5$. (c) STM  topographic image on the Sb surface of KV$_3$Sb$_5$ above $T_\mathrm{CDW}$ (upper) and below  $T_\mathrm{CDW}$ (down). Below $T_\mathrm{CDW}$, the surface shows a $2\times2$ modulation. (d) Fourier transform of the Sb topographic image above $T_\mathrm{CDW}$ (upper) and below  $T_\mathrm{CDW}$ (down)~\cite{Hasan2021NM}. The Fourier transform shows only Bragg peaks ($\mathbf{Q}_\mathrm{Bragg}$) at $T>T_\mathrm{CDW}$ and additional order peaks ($\mathbf{Q}_\mathrm{2\times2}$) appears at $T<T_\mathrm{CDW}$. (e) Schematic of SoD and TrH distortion within the Kagome layer in the CDW order. (f)  Momentum-dependent CDW gaps at the Fermi surface for KV$_3$Sb$_5$~\cite{ZhouXJ2022NC}. (g) Nematic phase revealed by STM topography and the corresponding Fourier transform in CsV$_3$Sb$_5$~\cite{ChenXH2022Nature}. (h)  Setup for angle-dependent resistivity measurement with the field and current applied within the ab-plane. (i) Corbino-shaped setup for angle-dependent c-axis resistivity measurement~\cite{WenHH2021NC}. (j) Temperature dependence of C2 symmetry for CsV$_3$Sb$_5$. The solid symbols denote the resistivity difference measured at 0.4 and 5 T and the open symbols denote the amplitude of C2 peak from Fourier transform. (k) Polar plot for angle-dependent c-axis resistivity below 2.4~T (left) at $T<T_\mathrm{c}$, showing two-fold symmetry, and above 2.4~T (right),  with the two-fold pattern rotated by 90$^\mathrm{o}$.  Here, (a-b) are taken from Ref.~\citenum{Wilson2020PRL}; (c-d) are taken from Ref.~\citenum{Hasan2021NM}; (f) is taken from Ref.~\citenum{ZhouXJ2022NC}; (g) is taken from Ref.~\citenum{ChenXH2022Nature}; (i-k) are taken from Ref.~\citenum{WenHH2021NC}.
	}
	\label{Fig3}
\end{figure*}

For CsV$_3$Sb$_5$, QO measurements have unveiled complex oscillation patterns. Fourier transform allows the resolution of up to 25 frequencies ($F$)~\cite{Welp2023PRL}. They are categorized into three regimes. In the low-$F$ regime (less than 250 T), four peaks of 18, 26, 73, and 92 T have been identified in various reports~\cite{Wilson2021PRX,ChenXH2021PRB}, as seen in {Fig~\ref{Fig2}a,b}. The estimated $m^*$ for these bands are relatively small, approximately one-tenth of the free electron mass. The peak of 73~T is consistently reported and attributed to a small Fermi pocket near the $L ~(\bar{M})$ point of the unfolded BZ~\cite{LeiHC2021PRL,Welp2023PRL} in {Fig~\ref{Fig1}f}. However, the other three $F$ are not well-resolved by ARPES and DFT calculations, necessitating further investigation. In the mid-$F$ regime ($250<F<2500$~T), a dozen of $F$ have been identified seen in {Fig.~\ref{Fig2}c,d}, though the exact number varies across different reports~\cite{Wilson2021PRX,Kwok2022PRB,Welp2023PRL}. This variation is due to differences in sample quality, the maximum of applied field and the resolution of QO techniques among different researches. $F$ in this regime is believed to account for the major Fermi pockets of CsV$_3$Sb$_5$. For instance, two $F$ around 736 and 804 T arise from the bands derived from Dirac crossings at the $K$ and $H$ points, respectively (see the triangles in {Fig~\ref{Fig2}e})~\cite{Welp2023PRL}. In the high-$F$ regime ($>2500$~T),  a few peaks have been identified, reaching up to 10 kT~\cite{Welp2023PRL}, as shown in the inset of {Fig~\ref{Fig2}d}. The estimated size of the corresponding Fermi surface is comparable to or even exceeds that of the folded first BZ in the CDW state. Chapai $et~al.$ attributed this observation to magnetic breakdown~\cite{Welp2023PRL}, a phenomenon arising from quantum tunneling of quasiparticles across closely spaced Fermi surfaces in the $k$-space.

The dimensionality of Fermi surface  can be explored by angle dependent QO measurements. For a 3D isotropic Fermi surface, $F$ is angle independent. While in 2D case, $F$ follows $F(\theta)=F(0^\circ)/\cos\theta$, where $\theta$ is the angle of $B$. In CsV$_3$Sb$_5$, the presence of both quasi-2D and 3D Fermi pockets at different $F$ has been demonstrated by previous studies~\cite{LeiHC2021PRL,Kwok2022PRB}. It reflects subtle features of the band structure induced by the specific organization of the CDW state. 
In addition to the frequency and amplitude, the phase ($\phi$) of QO also plays a pivotal role in uncovering non-trivial band topology. $\phi$ is expressed as: $\phi=2\pi(\frac{F}{B}+\frac{1}{2}-\frac{\Phi_\mathrm{B}}{2\pi}+\delta)$, where $\Phi_\mathrm{B}$ represents the Berry phase and $\delta$ is a dimentionality factor with $\delta=0$ for 2D and $\delta=\pm \frac{1}{8}$ for 3D Fermi surface. In {Fig.~\ref{Fig2}f}, the Landau fan diagram for CsV$_3$Sb$_5$ reveals non-trivial Berry phase ($\Phi_\mathrm{B}=\pi$) at $F\approx73$~T, underscoring the topological character of the corresponding band~\cite{LeiHC2021PRL,Welp2023PRL}. This serves an important indication of topologically nontrivial band structure, a feature frequently discussed in kagome systems.  Note that KV$_3$Sb$_5$ and RbV$_3$Sb$_5$ also exhibit rich QO spectra that bear recognizable similarity to CsV$_3$Sb$_5$~\cite{LeiHC2021CPL,Goh2023APL}. However, there is a slight difference in $F$, which may reflect subtle variations in the nature of CDW distortion in the three materials.

%Note that reports in this area are somewhat controversial. Some suggest that the majority of the bands display a quasi-2D nature~\cite{Goh2022PRB,Kwok2022PRB}, while others reveal a strong deviation from 2D nature in most bands~\cite{TianML2023Arxiv}. This ongoing debate highlights the need for further sophisticated investigations.
%,Goh2023JPM

~\\
\noindent \textit{3. CDW order.}
The CDW develops below $T_\mathrm{CDW}\approx78, 104$ and $94$~K in AV$_3$Sb$_5$ with A= K, Rb and Cs, respectively. {In Fig.~\ref{Fig3}a}, an anomaly in $T$-dependent resistivity ($\rho$) marks the CDW transition for CsV$_3$Sb$_5$. It corresponds to the sharp peak observed in heat capacity {in Fig.~\ref{Fig3}b}, implying a first-order transition~\cite{Wilson2020PRL}. 
The CDW modulation of local DOS has been extensively mapped using surface-sensitive scanning tunneling microscopy (STM) techniques~\cite{Zeljkovic2021Nature,Hasan2021NM,Hasan2021PRB,ChenXH2021PRX}, {as exemplified in Fig~\ref{Fig3}c}. Fourier transform, presented in {Fig.~\ref{Fig3}d}, reveals six additional ordering peaks, indicating a 3Q CDW order with a $2\times2$ superlattice modulation. This modulation corresponds to a breathing mode of kagome lattice showing either the Star of David (SoD) or trihexagonal (TrH) distortions~\cite{YanBH2021PRL}, depicted in {Fig~\ref{Fig3}e}. The TrH phase is predicted to be more energetically favorable~\cite{YanBH2021PRL}.
Moreover, STM indicates a CDW gap of approximately 50~meV ~\cite{Zeljkovic2021Nature,Hasan2021NM,ChenXH2021PRX,FengDL2021PRL}, which is strongly anisotropic over the V-derived Fermi surfaces, as evidenced by ARPES in {Fig.~\ref{Fig3}f}~\cite{ZhouXJ2022NC}. 

%(also referred to the inverse Star of David)
%,WangZW2021PRB

The mechanisms underlying the CDW state are complicated. In {Fig.~\ref{Fig3}f}, the anisotropic CDW gap shows a pronounced maximum around the M point~\cite{ZhouXJ2022NC}, implying the significance of inter-saddle-point scattering channels between different vHSs~\cite{Sato2021PRB}. This point was supported by other ARPES measurements~\cite{Comin2023NM,Borisenko202PRL} as well as several studies employing optical spectroscopy~\cite{WenHH2021PRB}. 
Moreover, hard X-ray scattering experiments on (Cs,Rb)V$_3$Sb$_5$ did not detected acoustic phonon anomalies, indicative of weak electron-phonon coupling~\cite{MiaoH2021PRX}. Taken together, these results suggest the dominant role of Fermi surface nesting in driving the CDW transition. 
In stark contrast, other investigations, including Raman scattering~\cite{XiXX2022npj}, neutron scattering~\cite{DaiPC2022PRB}, and optical conductivity~\cite{Tsirlin2022npj} measurements, observed strong electron-phonon coupling in AV$_3$Sb$_5$, emphasizing its importance in the development of the CDW state.  Evidence for this coupling is also seen in a kink in the energy dispersion observed in ARPES measurements of KV$_3$Sb$_5$~\cite{ZhouXJ2022NC}. DFT calculations revealed soft phonon modes at both M and L points of the BZ, and their condensation is proposed to facilitate the formation of the CDW~\cite{YanBH2021PRL,Harter2021PRM}. Furthermore, a combination of ARPES and DFT studies in CsTi$_3$Bi$_5$, a Ti-based kagome system analogous to AV$_3$Sb$_5$, reported that the presence of van Hove singularities alone is insufficient to drive the CDW transition~\cite{HeJF2023PRL}. It appears that both the Fermi surface nesting and electron-phonon coupling are essential for the CDW order.

%ZhaoZX2021Arxiv,Tsirlin2021PRB,Fernandes2021PRB,

According to DFT calculations~\cite{YanBH2021PRL}, phonon softening at $L$ point leads to a structural modulation along the c-axis in AV$_3$Sb$_5$. The 3D CDW order in this kagome family has been reported in various studies~\cite{Wilson2021PRX,ChenXH2021PRX,MiaoH2021PRX,Harter2021PRM}, with slight discrepancies among its members. 
High-resolution X-ray diffraction (XRD) measurements revealed a $2\times2\times2$ lattice distortion in KV$_3$Sb$_5$ and RbV$_3$Sb$_5$~\cite{Wilson2023PRM,MiaoH2021PRX}, characterized by a staggered trihexagonal (TrH) arrangement, i.e. a $\pi$ phase shift between neighboring kagome layers~\cite{Comin2023NM}. 
For CsV$_3$Sb$_5$, both $2\times2\times2$ and $2\times2\times4$ distortions were observed to coexist~\cite{Wilson2021PRX,Wilson2023PRM,PengYY2023PRR}, and one study reported a transition from the $2\times2\times4$ to the $2\times2\times2$ order as $T$ decreases below 60 K~\cite{Geck2023PRB}. 
The complex 3D CDW states in CsV$_3$Sb$_5$ are thought to result from a stacking of kagome sheets exhibiting alternating SoD and TrH distortions.~\cite{Comin2023NM,Wilson2023PRM}.
It appears that the different orders in AV$_3$Sb$_5$ are close in energy and highly sensitive to external perturbations such as disorder and frozen strains~\cite{Wilson2023PRM,PengYY2023PRR}.

%,Sanna2023PRR,ShiM2022PRB

~\\
\noindent\textit{Rotational symmetry breaking.}
Besides translation symmetry breaking, rotational symmetry breaking from C6 to C2 has been extensively documented in the CDW phase. Previous STM study revealed lattice  modulation along one of the three crystallographic directions, resulting in an imbalance among the amplitude of three CDW  vectors ($\textbf{Q}_\mathrm{2a_0}$)~\cite{ChenXH2022Nature}, as seen in {Fig.~\ref{Fig3}g}. This C2 symmetry appears at the Fermi level and is linked to the electronic nematic phase. This transition occurs at $T_\mathrm{nem}\approx~35$~K according to the measurements of nuclear magnetic resonance (NMR) and elastoresistance~\cite{ChenXH2022Nature}. Unidirectional electronic scattering of coherent quasiparticles was also detected by STM below 30~K~\cite{Zeljkovic2023NP}. Additionally, a 4$a_0$ stripe order was observed by some STM studies in CsV$_3$Sb$_5$ (at $T^*\approx60~K$)~\cite{Zeljkovic2021Nature,Zeljkovic2023NP} and RbV$_3$Sb$_5$~\cite{Hasan2021PRB}, but not in KV$_3$Sb$_5$~\cite{Hasan2021NM,Zeljkovic2022NP}. This order is thought to be a surface state, as it is not resolved by bulk-sensitive techniques like NMR and XRD.

The rotational symmetry breaking can be effectively resolved by angle-dependent magnetoresistance (AMR) measurements~\cite{ZhaoZX2021CPL,WenHH2021NC}. An early in-plane AMR study detected a twofold anisotropy, whose orientation rotates by 60$^\mathrm{o}$ upon entry into the SC phase~\cite{ZhaoZX2021CPL}. In this setup, with the current ($I$) and $B$ applied within the ab-plane (seen in { Fig.~\ref{Fig3}h}), orbital effects induced by Lorentz force generally lead to artificial C2 symmetry. Subsequently, Xiang $et~al.$ measured angle-dependent c-axis resistivity ($\rho_\mathrm{c}$) by using a Corbino-shaped electrode configuration, as illustrated in { Fig.~\ref{Fig3}i}, wherein $B$ rotates within the plane, always normal to $I$~\cite{WenHH2021NC}. This measurement revealed a twofold rotational symmetry below $40-60$~K, persisting into the SC state as shown in { Fig.~\ref{Fig3}j,k}. In the SC phase, the field direction, corresponding to the minimum resistance, is aligned with one of the crystalline axes, suggesting a maximum in the SC gap, as seen in { Fig.~\ref{Fig3}k}.  When a stronger field was applied to disrupt superconductivity, the twofold pattern rotated by 90$^\mathrm{o}$, highlighting the nematicity in the normal state. Furthermore, recent calorimetry measurements showed a twofold anisotropy superimposed on a sixfold background in the in-plane upper critical field, indicating a nematic SC order parameter~\cite{Yonezawa2024NC}. 

%\textbf{ChenJH2024NC}

\begin{figure*}[thb]
	\begin{center}
		\includegraphics[width=18cm]{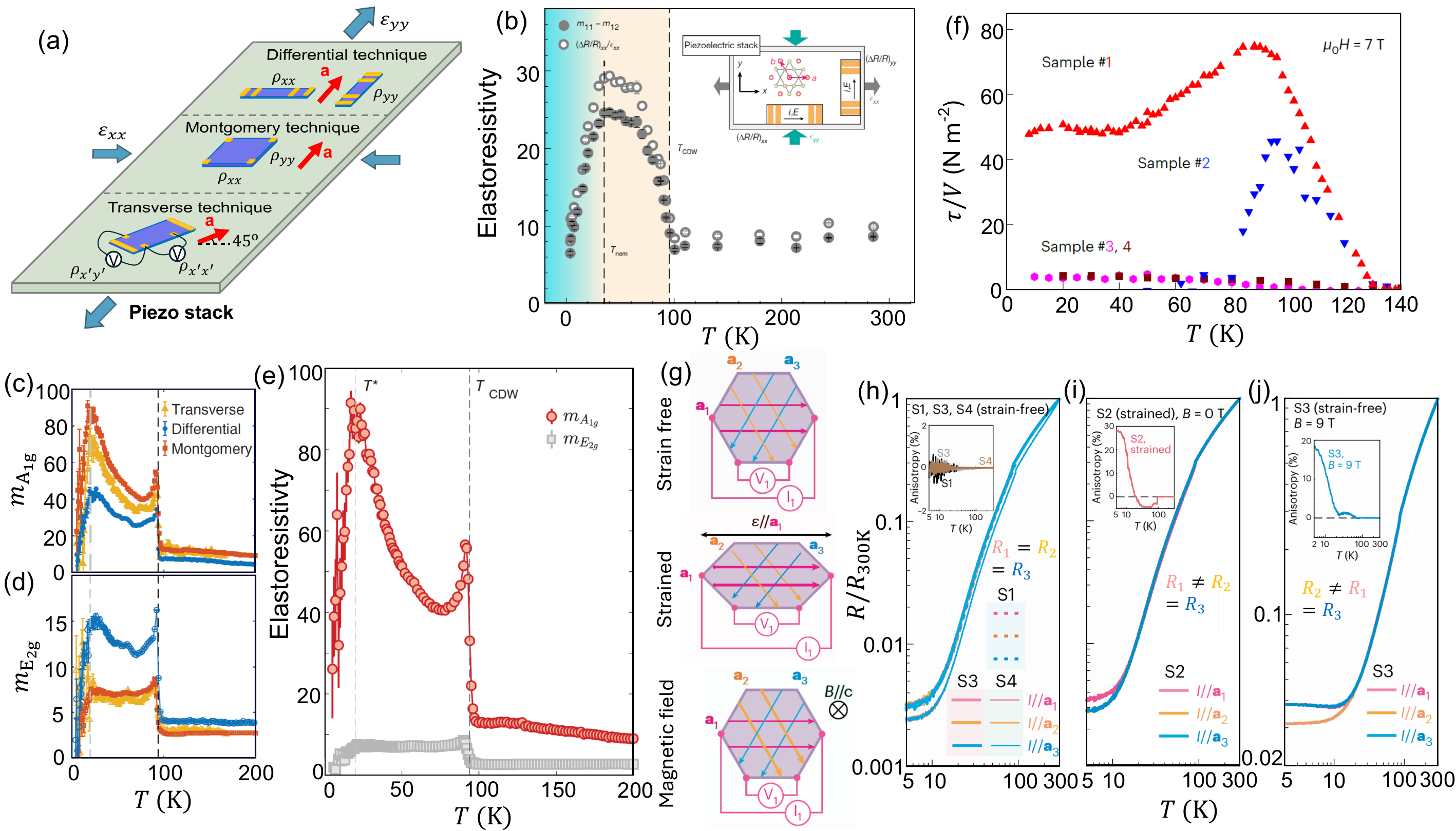}
	\end{center}
	\setlength{\abovecaptionskip}{-8 pt}
	\caption{(a) Schematic of three setups for elastoresistivity measurements: the differential technique, Montgomery technique and transverse technique. (b)  Temperature dependence of elastoresistivity coefficients ($m_{E_\mathrm{2g}}$) for CsV$_3$Sb$_5$ measured using the differential technique~\cite{ChenXH2022Nature}. (c-d) Temperature dependence of elastoresistivity coefficients in the isotropic $A_\mathrm{1g}$ and anisotropic $E_\mathrm{2g}$ channels for CsV$_3$Sb$_5$, measured with all three techniques, respectively~\cite{ChuJH2024PRX}. (e) Comparison of $m_{A_\mathrm{1g}}$ and  $m_{E_\mathrm{2g}}$ from the data in (c-d), obtained using the Montgomery technique. (f) Temperature dependence of magnetic torque normalized by the sample volume for different samples of CsV$_3$Sb$_5$~\cite{Matsuda2024NP}. (g) Schematic illustration of in-plane electric transport under different conditions: strain free, in-plane strain and perpendicular external field. The resistance is measured along the three main axes. (h-j) Temperature dependence of tri-directional resistance for CsV$_3$Sb$_5$ under conditions of strain-free (h), strain (i) and magnetic field (j)~\cite{Moll2024NP}.  Here, (b) is taken from Ref.~\citenum{ChenXH2022Nature}; (c-e) are taken from Ref.~\citenum{ChuJH2024PRX}; (f) is taken from Ref.~\citenum{Matsuda2024NP}; (g-j) are taken from Ref.~\citenum{Moll2024NP}.
	}
	\label{Fig4}
\end{figure*}

Elastoresistivity measurements are an effective tool for probing nematic phase transitions. Elastoresistivity, a fourth-rank symmetric tensor, quantifies the change in resistivity under external strain, expressed as

\begin{equation}
	m_\mathrm{ij}=\frac{\partial(\Delta\rho/\rho)_\mathrm{i}}{\partial\varepsilon_\mathrm{j}}
\end{equation}

\noindent
where $\varepsilon_{j}$ represents the strain tensor, and $\Delta \rho$ denotes the resistivity change induced by strain.  The indices $i, j=1-6$ follow Voigt notation with $1=xx$, $2=yy$, $3=zz$, $4=yz$, $5=zx$ and $6=xy$. Elastoresistivity coefficients can be grouped into different symmetry channels according to the irreducible representations (irrep) of the lattice point group. For the Kagome system with $D_\mathrm{6h}$ symmetry, the coefficient of the isotropic $A_\mathrm{1g}$ irrep defined as $m_{A_\mathrm{1g}}=m_{11}+m_{12}+m_{13}[2\nu_\mathrm{ac}/(1-\nu_\mathrm{ab})]$ with $\nu_\mathrm{ac}$ and $\nu_\mathrm{ab}$ the Poisson ratio.  For the anisotropic $E_\mathrm{2g}$  irrep, the coefficient is $m_{E_\mathrm{2g}}=m_{11}-m_{12}$. The measurement of elastoresistivity generally employs three methods: the differential elastoresistivity technique, the modified Montgomery technique, and the transverse technique~\cite{ChuJH2024PRX}, as illustrated in {Fig.~\ref{Fig4}a}.

In the context of nematic phase transitions, nematic susceptibility, linked to $m_{E_\mathrm{2g}}$, is expected to diverge as $T$ approaches $T_\mathrm{nem}$. In {Fig.~\ref{Fig4}b}, the $T$-dependence of $m_{E_\mathrm{2g}}$, measured by differential techniques, shows a sharp jump at $T_\mathrm{CDW}$ and continues to rise down to $T_\mathrm{nem}\approx35$~K for CsV$_3$Sb$_5$~\cite{ChenXH2022Nature}. This phenomenon is distinct from the Curie-Weiss trend typically observed in a continuous nematic phase transition. The authors attributed it to nematic fluctuations pinned by the pre-existing C2 structural distortion in the 3D CDW phase. A subsequent work reported optimized superconductivity in the vicinity of a nematic quantum critical point in Ti-doped CsV$_3$Sb$_5$~\cite{Kim2023NC}. 

%the $E_{2g}$  elastoresistivity coefficient

Conversely, Liu $et~al.$ carried out a more systematic study of the elastoresistivity using all three techniques and observed a distinct behavior in {Fig.~\ref{Fig4}c-e}~\cite{ChuJH2024PRX}. Specifically, $m_{A_\mathrm{1g}}$, rather than $m_{E_\mathrm{2g}}$, show a prompt jump at $T_\mathrm{CDW}$, followed by a continuous increase as $T$ decreases to 20~K in {Fig.~\ref{Fig4}e}.  In contrast, $m_{E_\mathrm{2g}}$ shows a slight jump but remains flat both below and above $T_\mathrm{CDW}$. The authors suggested that the pronounced $m_{E_\mathrm{2g}}$ signal in previous reports arises from cross-contamination from the  $A_\mathrm{1g}$ channel, a common  problem in differential techniques. Their findings indicate the absence of nematic phase transition within the CDW phase in CsV$_3$Sb$_5$. Similar observations were reported in Ref.~\citenum{Hardy2024PRL}. Several reports proposed that the C2 symmetry arises from the interlayer coupling, i.e. a $\pi$-phase shift in the stacking of adjacent SoD or TrH distorted kagome layers~\cite{ChenXH2022Nature,ChenXH2021PRX,WenHH2021NC,Hardy2024PRL,YanBH2021PRB,Balents2021PRB}.

In this context, recent magnetic torque ($\tau$) measurements uncovered a two-fold in-plane magnetic anisotrpy with an onset temperature of around 130~K, notably above $T_\mathrm{CDW}$, as shown in {Fig.~\ref{Fig4}f}~\cite{Matsuda2024NP}. The hysteretic behavior in $\tau$ under a conically rotated magnetic field provides thermodynamic evidence supporting the emergence of an odd-parity nematic order, a phenomenon that cannot be detected by an even-parity probe such as elastoresistivity. However, other bulk thermodynamic probes, such as elastocaloric effect~\cite{ChuJH2024PRX} and thermal expansion~\cite{Hardy2024PRL}, didn't observe any anomaly above $T_\mathrm{CDW}$.     

A unified understanding of the C2 symmetry and nematicity in kagome superconductors remains an important ​target. Guo $et~al.$ reported the absence of anisotropy in strain-free CsV$_3$Sb$_5$ samples over the full temperature range, seen in {Fig.~\ref{Fig4}g-j}~\cite{Moll2024NP}. However, once either a weak perpendicular field or in-plane strain are present, a pronounced in-plane transport anisotropy appears below $T_\mathrm{CDW}$ in {Fig.~\ref{Fig4}h-j}. While both perturbations yield similar anisotropic behavior, their underlying mechanisms differ: strain directly breaks rotational symmetry, whereas the field is proposed to stabilize the flux order coexisting with bond order~\cite{Moll2024NP}. This stark contrast in origin highlights the system's sensitivity to external perturbations, which may offer some insights to the controversy surrounding the entangled CDW orders.

\begin{figure*}[thb]
	\begin{center}
		\includegraphics[width=18cm]{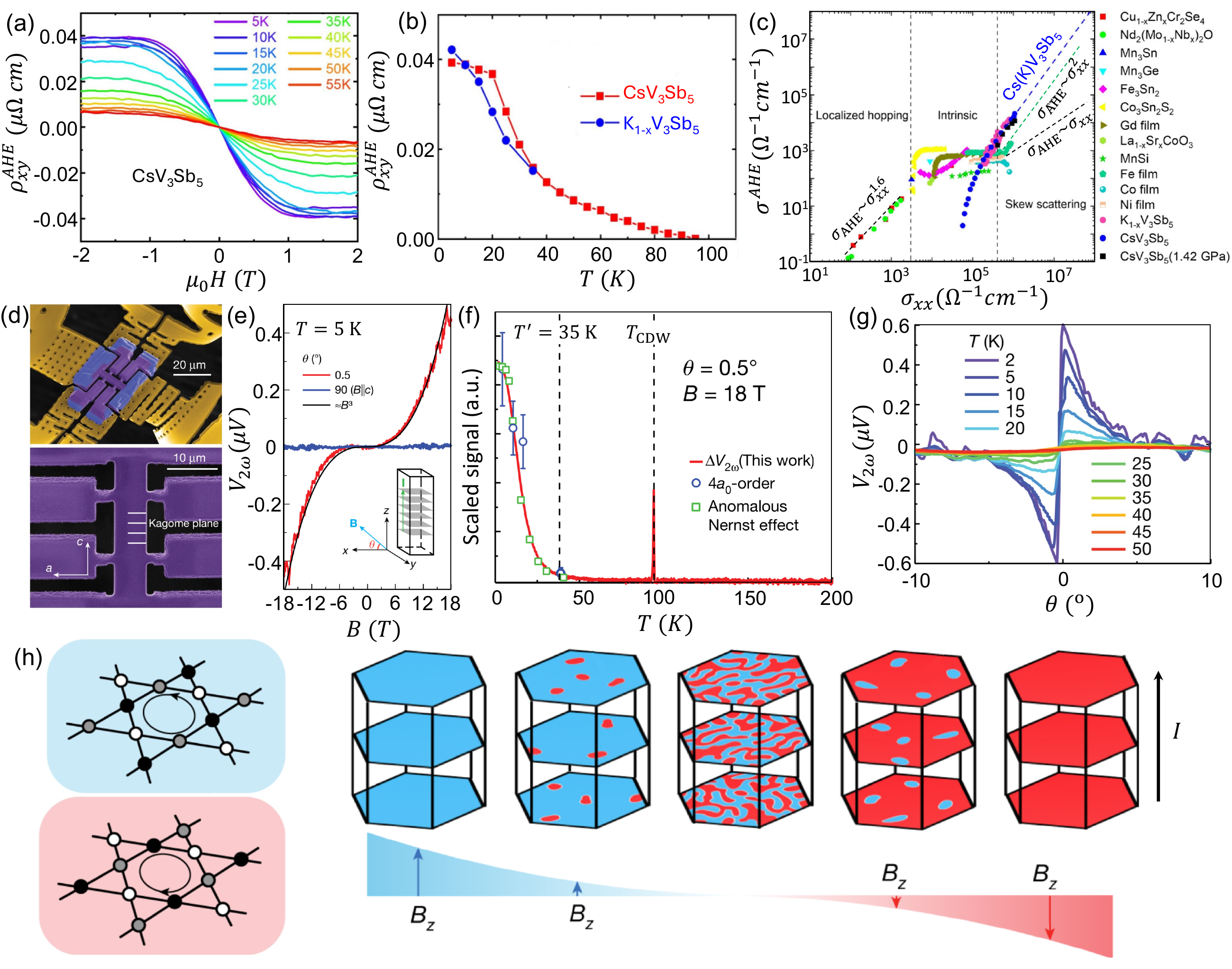}
	\end{center}
	\setlength{\abovecaptionskip}{-3 pt}
	\caption{(a) Anomalous Hall resistivity ($\rho_\mathrm{xy}^\mathrm{AHE}$) measured at various $T$ below $T_\mathrm{CDW}$ in CsV$_3$Sb$_5$~\cite{ChenXH2021PRB}. (b) Temperature dependence of $\rho_\mathrm{xy}^\mathrm{AHE}$ for CsV$_3$Sb$_5$ and KV$_3$Sb$_5$, showing an increase below $T_\mathrm{CDW}$~\cite{ChenXH2021PRB}. (c) Scaling of anomalous Hall conductivity ($\sigma^\mathrm{AHE}$) with electric conductivity ($\sigma_\mathrm{xx}$) for multiple materials including CsV$_3$Sb$_5$~\cite{ChenXH2021PRB}. (d) Low-strain device fabricated by focused ion beam, with current applied along the c-axis. (e) Field dependence of second harmonic voltage ($V_\mathrm{2\omega}$) with current applied along the c-axis. (f) Temperature dependence of $V_\mathrm{2\omega}$, compared with signals from other probes.   $V_\mathrm{2\omega}$ shows a peak at $T=T_\mathrm{CDW}$ and significantly increases at $T<35~K$. (g) Angular dependence of $V_\mathrm{2\omega}$ at $B=18$~T and various temperatures. (h) Schematic illustration of chiral charge order domains with chirality controlled by the out-of-plane field component~\cite{Moll2022Nature}.   Here, (a-c) are taken from Ref.~\citenum{ChenXH2021PRB}; (d-h) are taken from Ref.~\citenum{Moll2022Nature}.
	}
	\label{Fig5}
\end{figure*}

~\\
\noindent \textit{Time reversal symmetry breaking.}
The most outstanding feature of the CDW order in AV$_3$Sb$_5$ is TRS-breaking. Although the absence of long-range magnetic ordering is confirmed by neutron scattering experiments~\cite{Toberer2019PRM}, zero-field $\mu$SR measurements detected an apparent increase of the internal field below $T_\mathrm{CDW}$, which persists into the SC phase, suggesting the presence of unconventional TRS-breaking charge orders~\cite{Guguchia2022Nature}. TRS-breaking signals were also observed through magneto-optical Kerr effect and circular dichroism measurements, with the onset coinciding with the CDW transition~\cite{WuL2022NP}. Simultaneously, scanning birefringence measurements revealed three C2 domains, approximately 100~$\mu m$ in size, oriented at $2\pi/3$ angles with respect to each other~\cite{WuL2022NP}. From the STM measurements, the intensity peaks of charge modulation in the Fourier transform show chiral anisotropy, the chirality of which can be reversed by applying an external magnetic field in the opposite direction~\cite{Hasan2021NM, Hasan2021PRB}. Moreover, recent studies revealed the chirality of CDW peaks modulated by linearly polarized light along high-symmetry directions~\cite{Madhavan2024Nature}. This optically switchable chirality, combined with field tunability, suggests a peculiar piezo-magnetic effect. Such behavior strongly indicates TRS-breaking, potentially arising from orbital magnetization inherent to the CDW phase.

%,Luetkens2022PRR,WangZW2021PRB,WangNL2022PRB

Transport measurements are an importance probe for detecting TRS-breaking phases. A giant anomalous Hall (AHE) effect has been reported in  AV$_3$Sb$_5$ with its onset temperature concurrent with $T_\mathrm{CDW}$~\cite{Ali2020SA,ChenXH2021PRB}, as shown in {Fig.~\ref{Fig5}a-b}. 
The scaling of $\sigma_\mathrm{AHE}$ and $\sigma_\mathrm{xx}$ for AV$_3$Sb$_5$ is presented in {Fig.~\ref{Fig5}c}~\cite{Ali2020SA}, in comparison with other systems. The scaling relationship differs across different regimes as $\sigma_\mathrm{xx}$ increases. In the localized hopping regime, a side-jump effect causes $\sigma_\mathrm{AHE}$ to follow a $\sigma_\mathrm{xx}^{1.6}$ dependence. In the intrinsic regime, the AHE is generated by the Berry curvature, making $\sigma_\mathrm{AHE}$ independent of $\sigma_\mathrm{xx}$. Finally, in the pure regime, skew scattering dominates and $\sigma_\mathrm{AHE}$ becomes linear in $\sigma_\mathrm{xx}$. In {Fig.~\ref{Fig5}c}, $\sigma_\mathrm{AHE}$ of AV$_3$Sb$_5$ follows a quadratic scaling, similar to that of iron in the skew scattering regime, but with a much larger skew constant. While the exact mechanism for the giant AHE in AV$_3$Sb$_5$ is yet to be resolved, it is thought to be closely tied to the enhanced skew scattering in the CDW state and the large Berry curvature arising from the kagome lattice~\cite{Ali2020SA,ChenXH2021PRB}. Besides AHE, the exotic feature could also be manifested in anomalous effects in thermal and thermal-electric transports~\cite{HeMQ2023Tungsten,LiL2024NC}. 
It is important to note that the AHE in AV$_3$Sb$_5$ does not exhibits a typical hysteresis behavior as that of ferromagnetic ordering, therefore it is not definitive proof of TRS-breaking. 

%LuoJL2022PRB，Felser2022PRB

%The anomalous Hall conductivity ($\sigma_\mathrm{AHE}$) in CsV$_3$Sb$_5$ achieves $2.1\times10^4~\Omega^{-1}\mathrm{cm}^{-1}$, larger than those in most of the ferromagnetic metals~\cite{ChenXH2021PRB}.  such as anomalous Nernst~\cite{HeMQ2021PRB,Felser2022PRB,LuoJL2022PRB} and  anomalous thermal Hall effect~\cite{LiL2024NC}. 
 
Using focused ion beams, Guo $et~al.$ fabricated strain-relaxed CsV$_3$Sb$_5$ devices with the current applied along the c-axis and reported field-switchable nonlinear transports, as seen in {Fig.~\ref{Fig5}d-e}. For $B$ applied at small $\theta$ with respect to the ab-plane,  they observed second harmonic voltage generation ($V_\mathrm{2\omega}$) at low temperatures, closely linked to the transitions within the CDW state, as shown in {Fig.~\ref{Fig5}f}. 
This phenomenon could be interpreted by the electronic magnetochiral anisotropy (eMChA),  which requires orthogonal orientations of the electric current $\mathbf{I}$, magnetic field $\mathbf{B}$ and the direction of inversion symmetry breaking.
%This phenomenon could be interpreted by the electronic magnetochiral anomaly (eMChA), expressed as $V(\mathbf{B,I})=IR_0(1+\gamma(\mathbf{P}\times\mathbf{B})\cdot\mathbf{I})$, which requires orthogonal orientations of $\mathbf{I}$, $\mathbf{B}$ and $\mathbf{P}$ (direction of inversion symmetry breaking).
In  {Fig.~\ref{Fig5}g}, $V_\mathrm{2\omega}$ is zero at $\theta=0^\mathrm{o}$ ($\mathbf{B}\parallel$ ab-plane) and increases rapidly as $\theta$ deviates from zero. It peaks at $\theta\approx0.5^\mathrm{o}$, before diminishing at larger $\theta$ (larger out-of-plane component of field $B_\mathrm{z}$). Moreover, the polarity of $V_\mathrm{2\omega}$ is reversed as $\theta$ evolves from positive to negative by inverting $B_\mathrm{z}$. Such a field-switching polarity suggests the TRS-breaking charge order. 

The microscopic origin of the observed second harmonic generation remains unclear. The authors proposed a phenomenological model involving chiral charge order domains, as illustrated in  {Fig.~\ref{Fig5}h}. The chiral domain structure is by nature spatially asymmetric, disrupting in-plane mirror symmetry and can be tuned by $B_\mathrm{z}$.  For small $\theta$, $B_\mathrm{z}$ induces an imbalance in domians of opposite chirality, leading to strong chiral scattering at domain walls and resulting in large eMChA. As $B_\mathrm{z}$ increases at larger $\theta$, the domains become fully polarized. In this case, the intrinsic chiral scattering process within the domains is too weak to produce observable eMChA. At $\theta=0^\mathrm{o}$, domains of opposite chirality are in equilibrium. Therefore, the global transport measurements observed averaged signals with vanishing eMChA. In this context, $B_\mathrm{z}$ adjusts domain chirality, while the in-plane component ($B_\mathrm{in}$) ensures orthogonality, enabling finite eMChA.

Theoretically, the TRS-breaking charge order in AV$_3$Sb$_5$ is likely a chiral flux phase, a long-sought order first proposed for the honeycomb lattice in Haldane's model~\cite{Haldane1988PRL} and later for the square lattice of cuprates by Varma~\cite{Varma1997PRB}. In the Kagome lattice,  the chiral flux phase with orbital current is energetically favorable near the van Hove filling due to the unique sublattice structure~\cite{HuJP2021SB,Neupert2021PRL}. The imbalance of anti-clockwise triangle current flux loop and a clockwise hexagonal flux loop leads to a net signal of TRS-breaking. 
However, it should be noted that a definitive conclusion regarding the chiral flux phase in AV$_3$Sb$_5$ remains controversial. Other conflicting studies, including $\mu$SR~\cite{Graf2021JPCM}, STM~\cite{Zeljkovic2022NP,WenHH2022PRB}, and polar optical Kerr effect~\cite{Kapitulnik2023PRL} measurements, have failed to detect signals of TRS-breaking. 

%Rahul2021PRB

\begin{figure*}[thb]
	\begin{center}
		\includegraphics[width=18cm]{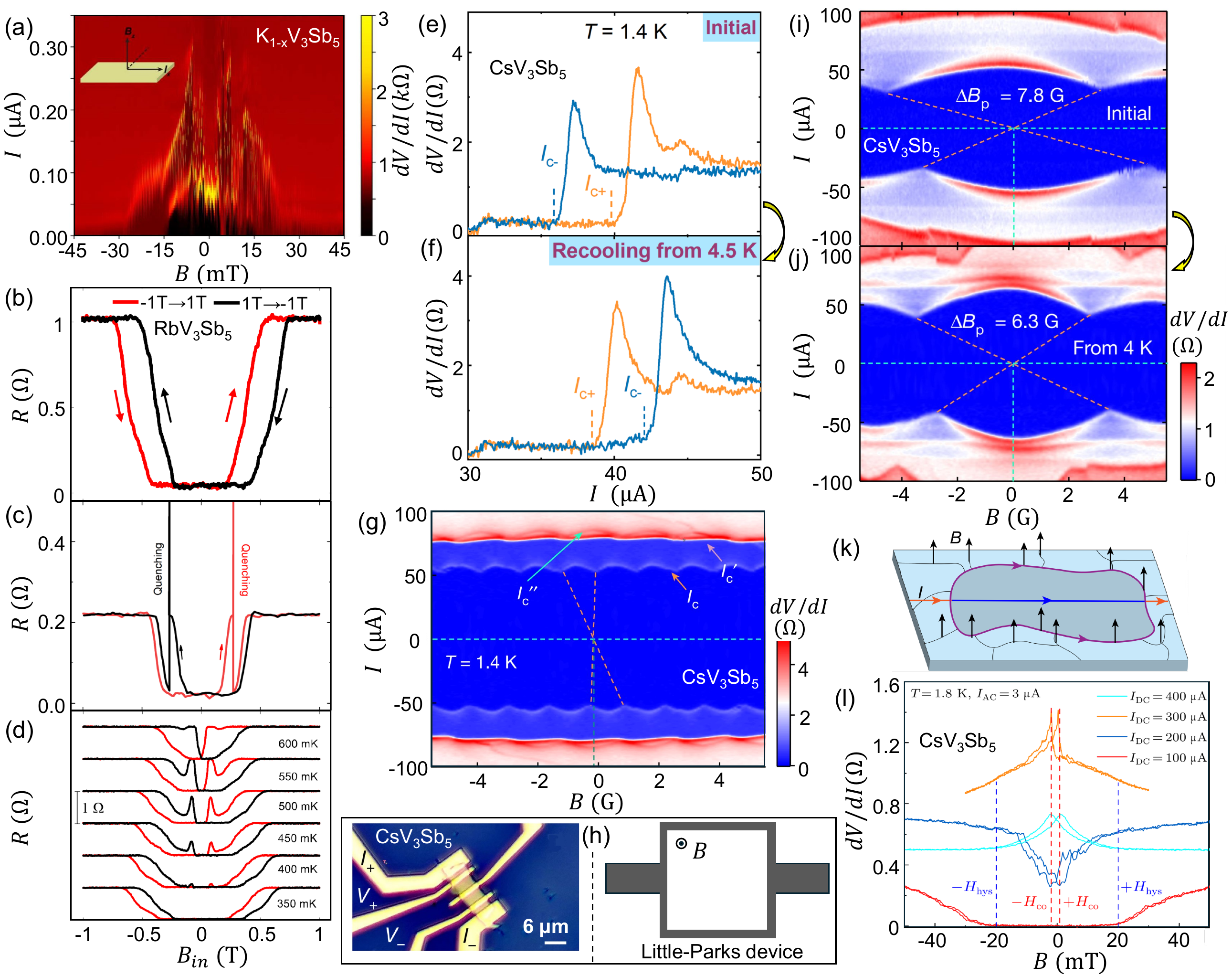}
	\end{center}
	\setlength{\abovecaptionskip}{-3 pt}
	\caption{(a) Color plots of $\mathrm{d}V/\mathrm{d}I$ versus current and out-of-plane field for K$_\mathrm{1-x}$V$_3$Sb$_5$-based Josephson junctions measured at 20~mK~\cite{Ali2023SA}. (b-d) Field hysteresis of magnetoresistance for RbV$_3$Sb$_5$ measured in the SC state by applying in-plane field~\cite{YuDP2024Arxiv}. (e-f) Zero-field SDEs for CsV$_3$Sb$_5$ with its polarity modulated by thermal cycling. (g) Color plots of $\mathrm{d}V/\mathrm{d}I$ versus current and out-of-plane field for  CsV$_3$Sb$_5$, where the critical current shows periodic oscillations. (h) Photo image of exfoliated CsV$_3$Sb$_5$ devices, compared with the configuration of a standard Little-Parks device. (i-j) Thermal modulation of oscillation patterns, showing variations in oscillation period. (k) Sketch of domain network, composed of chiral SC domains.~\cite{LeT2024Nature} (l) Field hysteresis of $\mathrm{d}V/\mathrm{d}I$ in the SC mixed state for CsV$_3$Sb$_5$ measured with out-of-plane field~\cite{LeT2024CPB}.   Here, (a) is taken from Ref.~\citenum{Ali2023SA}; (b-d) are taken from Ref.~\citenum{YuDP2024Arxiv}; (g-k) are taken from Ref.~\citenum{LeT2024Nature}; (l) is taken from Ref.~\citenum{LeT2024CPB}.
	}
	\label{Fig6}
\end{figure*}

~\\
\noindent\textit{4. Superconducting orders.}
The superconductivity of the Kagome systems remains a critical topic of research, with ongoing debates on the pairing mechanism and pairing symmetry. The interplay between the complex CDW order and superconductivity~\cite{ChenXH2022nature1,ChengJG2021PRL} leads to the PDW order~\cite{GaoHJ2021Nature,YinJX2024Nature} and further complicates the understanding of the SC state, raising the fundamental question of whether it is conventional or unconventional.

From the experimental viewpoint, thermal conductivity measurements in CsV$_3$Sb$_5$ revealed a residual linear term in $\kappa$ at zero-$T$~\cite{LiSY2024CPL}, which is often associated with nodal superconductivity. Similar results were observed in KV$_3$Sb$_5$ and RbV$_3$Sb$_5$ through $\mu$SR, where the SC penetration depth ($\lambda$) exhibited a linear temperature dependence~\cite{Khasanov2023NC}. In contrast, in CsV$_3$Sb$_5$, $\lambda$ measured by $\mu$SR~\cite{LeiHC2022npj} and TDO~\cite{Yuan2021SCPMA} followed an exponential temperature dependence, suggesting a fully gapped superconductivity. STM studies yielded conflicting results: both V-shaped (nodal gap) and U-shaped (nodeless gap) scanning tunneling spectroscopy (STS) profiles were observed in CsV$_3$Sb$_5$, suggesting the complexity of the gap  structure~\cite{FengDL2021PRL,GaoHJ2021Nature}.

Moreover, nuclear quadrupole resonance (NQR) measurements of the  121/123 Sb atoms showed a Hebel-Slichter coherence peak at $T_\mathrm{c}$ of CsV$_3$Sb$_5$, which is widely regarded as a hallmark for conventional $s$-wave superconductivity~\cite{LuoJL2021CPL}. Additionally, $T_\mathrm{c}$ in CsV$_3$Sb$_5$ is weakly sensitive to non-magnetic impurities introduced by electron irradiation, another indication of fully gapped pairing symmetry. 
By increasing the amount of impurities, $\lambda$, measured by TDO, reveals a gradual evolution from a highly anisotropic fully gapped SC state to an isotropic fully gapped state~\cite{Shibauchi2023NC}. Recent high-revolution ARPES measurements examined the SC gap in momentum space and revealed multiple gaps with a highly anisotropic gap structure (80\% anisotropy) at the hexagonal Fermi surface near the Brillouin zone boundary and isotropic gaps at other regions~\cite{Okazaki2024Arxiv}. Similar to this observation, STM studies unveiled residual in-gap states within the SC gaps and the corresponding Fermi arcs in $k$-space. These features might be related to the specific orbital structure of the PDW order~\cite{YinJX2024Nature}. The multiple, anisotropic, fully-gapped, spin-singlet pair symmetry could help to reconcile the aforementioned conflicting observations.

On the other hand, first-principles calculations suggest that the electron-phonon coupling is too weak to account for the high $T_\mathrm{c}$ in AV$_3$Sb$_5$, pointing to the possibility of unconventional pairing~\cite{YanBH2021PRL}. However, it is also proposed that the on-site correlations may be weak, disfavor unconventional superconductivity. Instead, non-local electron correlations, such as nearest-neighbor and next-nearest-neighbor interactions, are thought to play a crucial role in these materials. Several theoretical studies incorporating both on-site $U$ and nearest-neighbor $V_1$ have predicted a rich phase diagram, spanning both conventional ($s$-wave) and unconventional (chiral $d\pm id$ or $f$-wave) superconductivity~\cite{LiJX2012PRB, WangQH2013PRB, Thomale2013PRL, Thomale2021PRL}.

In AV$_3$Sb$_5$, superconductivity develops within the pre-existing CDW states. The mutual evolution between these two orders tuned by hydrostatic pressure or chemical doping has been widely reported in different studies, with CsV$_3$Sb$_5$ displaying an abnormal double-dome SC phase diagram~\cite{ChenXH2022nature1,ChengJG2021PRL,Yuan2024SST}. It indicates a complicatd competition between the SC and CDW orders. Moreover, the ratio between $T_\mathrm{c}$ and $\lambda_\mathrm{eff}^{-2}$ is significantly higher than that in conventional superconductors, indicative of a dilute superfluid density~\cite{Guguchia2022Nature,Khasanov2023NC,LeiHC2022npj}. These characteristics collectively suggest unconventionality in the SC phase. 

%ChenJGPRR2021,Yuan2022PRB,GaoHJ2022SB

Another question is whether the SC phase, intervening with the TRS-breaking charge order, disrupts TRS. While given the pre-existence of TRS-breaking in the normal state, zero-$B$ $\mu$SR failed to detect an additional transition in the muon relaxation rate at $T_\mathrm{c}$~\cite{Guguchia2022Nature}. Subsequent experiments observed a slight upturn in the muon relaxation rate across $T_\mathrm{c}$, when the charge order is completely suppressed in AV$_3$Sb$_5$~\cite{Khasanov2023NC,YinJX2024NM}, which might be a signature of TRS-breaking superconductivity.
Furthermore, Deng $et~al.$ reported $B$-tunable chirality of PDW intensity peaks in STM, pointing to the chiral PDW nature in AV$_3$Sb$_5$~\cite{YinJX2024Nature}. When the CDW order is suppressed by Ta doping in Cs(V$_\mathrm{1-x}$Ta$_\mathrm{x}$)$_3$Sb$_5$, the same group observed a time-reversal asymmetric interference of Bogoliubov quasiparticles via external $B$ inversion, supporting TRS-breaking superconductivity, in agreement with $\mu$SR data~\cite{YinJX2024NM}. 

%Khasanov2022CP

Transport measurements have played a vital role in elucidating the pairing symmetry in AV$_3$Sb$_5$. Fitting $\lambda$, estimated from the self-field critical current ($I_\mathrm{c}$), suggests conventional $s$-wave superconductivity with strong coupling in CsV$_3$Sb$_5$~\cite{Goh2023NL}. However, Josephson junctions based on non-SC K$_\mathrm{1-x}$V$_3$Sb$_5$ show anomalous patterns in {Fig~\ref{Fig6}a}, including a minimum near zero $B$, which was interpreted by a possible spin-triplet pairing~\cite{Ali2023SA}. More recently, Wang $et~al.$ reported an unusual ``advanced'' hysteresis of the magnetoresistance in the SC state of RbV$_3$Sb$_5$, as presented in {Fig.~\ref{Fig6}b}~\cite{YuDP2024Arxiv}. The hysteresis is pronounced upon sweeping in-plane $B_\mathrm{in}$, but diminishes significantly and eventually disappears when $B$ is tilted away from the ab-plane.  
The finite-resistance state within the hysteresis loop appears metastable and can be quenched to zero resistance after applying large $I$ (heating process), as seen in {Fig.~\ref{Fig6}c}.  Additionally, the magneto-resistance exhibits reentrant superconductivity upon sweeping  $B_\mathrm{in}$,  as illustrated in  {Fig.~\ref{Fig6}d}. To explain these observations, the authors proposed a TRS-breaking spin-triplet p-wave pairing state, hosting net spin-magnetic moments. TRS-breaking SC domains, composed of reversed spin-polarized paring orders, interact with $B_\mathrm{in}$, giving rise to the observed phenomena. However, this interpretation does not clarify why the effects are exclusively influenced by $B_\mathrm{in}$. The proposed spin-triplet pairing contrasts sharply with prior reports~\cite{LeiHC2022npj,Yuan2021SCPMA,LuoJL2021CPL,Shibauchi2023NC,Okazaki2024Arxiv}.

SC diode effects (SDEs)  arise in systems lacking both TRS and inversion symmetry, manifesting as asymmetry in $I_\mathrm{c}$ with respect to the direction of current flow. Recently, SDEs have emerged as a novel tool for probing exotic SC orders in quantum materials. 
Le $et~al.$ reported  the observation of SDEs in CsV$_3$Sb$_5$ flakes under zero $B$ in {Fig.~\ref{Fig6}e}, which unveils internal TRS-breaking~\cite{LeT2024Nature}. The SDE polarity could be modulated by thermal cycling from $T$ slightly above $T_\mathrm{c}$ as seen in {Fig.~\ref{Fig6}f}, potentially indicating a dynamic domain structure associated with the SC order. Furthermore, the authors observed double-slit interference patterns (DSIP) in $I_\mathrm{c}$ upon applying a small external $B$, as shown in  {Fig.~\ref{Fig6}g}. These patterns resemble those seen in Little-Parks effects, which typically require a current loop with a hollow core, contrasting with the configuration of mechanically exfoliated CsV$_3$Sb$_5$ devices in {Fig.~\ref{Fig6}h}. The oscillation period, on the order of several Gauss, corresponds to a flux-penetration area (loop size) of several $\mu$m$^2$. The features of oscillations, such as the period and phase, also exhibit thermal modulation as seen in {Fig.~\ref{Fig6}i,j}. These phenomena are interpreted within a domain-network model, composed of TRS-breaking SC domains of opposite chirality, as illustrated in  {Fig.~\ref{Fig6}k}. 
The evolution of chiral SC order parameter from one domain to its neighbors of opposite chirality leads to the suppression of the order parameter and generates boundary supercurrents flowing along the domain walls.  This behavior mimics the supercurrent loop in a standard Little-Parks device, which is responsible for the observation of DSIP. 
The dynamics of SC domains can also be reflected by the observation of field hysteresis in the differential resistance within the mixed state, where vortices are depinned, seen in  {Fig.~\ref{Fig6}l}~\cite{LeT2024CPB}.
Taken together, the study of thermally modulated zero-field SDE and DSIP unveils dynamic TRS-breaking SC domains with edge supercurrents in CsV$_3$Sb$_5$. This approach may serve as an effective tool to explore potential TRS-breaking superconductors.

\noindent\textit{5. Summary and outlook.}
The kagome AV$_3$Sb$_5$ system has emerged as one of the most rapidly advancing subjects in condensed matter physics since its discovery in 2019. A number of exotic and intervened orders have been observed, yet several fundamental questions lack comprehensive interpretation. 
In the normal state, the system exhibits a plethora of electronic orders that are nearly degenerate in energy and highly sensitive to external perturbations such as strain, magnetic field and polarized optical excitation. A solid demonstration of the long-sought chiral flux order remains a crucial goal in this field and demands the cooperation of multiple experimental techniques. Recent experimental studies suggest that the orbital magnetism manifests as a piezomagnetic effect~\cite{Madhavan2024Nature,Moll2024NP}, which warrants further exploration. It's  generally expected that the multiple orders would lead to complex domain structures in AV$_3$Sb$_5$. C2 symmetry domains have been observed by polarized MOKE measurements~\cite{WuL2022NP}. While the detection of chiral domains with edge orbital current, an essential signature of chiral flux order, remains be a big challenge. This will hopefuly be solved by advanced field-sensitive scanning probes. 

%{\color{red}To resolve this issue, a comprehensive mapping of orbital current domains using spatially-resolved scanning probes is necessary, but such measurements are still lacking.}

In AV$_3$Sb$_5$, the SC order has converged to multiple, highly anisotropic, sign preserved, spin-singlet pair symmetry. 
When the CDW order is completely suppressed by chemical doping, an isotropic, fully gapped order parameter was resolved by ARPES, pointing to $s$-wave superconductivity~\cite{Okazaki2023Nature}. However, this conclusion disagrees with the observation of TRS-breaking signals in the SC state~\cite{LeT2024Nature,LeT2024CPB,YuDP2024Arxiv,YinJX2024NM}. A chiral $d+id$ state, combining nodeless superconductivity and broken TRS, may reconcile these conflicting observations. And the kagome sublattice symmetry potentially mitigates disorder-induced pairing breaking in spin-singlet $d$-wave states~\cite{Andersen2023PRB}.  A comprehensive understanding of the pairing symmetry will benefit from phase-sensitive techniques like Josephson junction experiments in the future. Moreover, Ge $et~al.$ identified exotic charge-4$e$ and charge-6$e$ superconductivity in the fluctuating region of Cooper pair condensate in CsV$_3$Sb$_5$~\cite{WangJ2024PRX}. Before drawing a firm conclusion, further studies are required to verify this unusual paring, particularly direct detection of fractional flux quanta by spatially resolved probes such as STM and scanning SQUID. 

Beyond the V-based Kagome system, novel AM$_3$X$_5$ variants, such as CsTi$_3$Bi$_5$ and CsCr$_3$Sb$_5$, have recently been discovered to exhibit superconductivity and intriguing correlated orders~\cite{GHJ2024NC,CaoGH2024Nature}.  These systems offer new opportunities to explore unconventional electronic phenomena, extending the frontiers of kagome physics.  A deeper understanding of the V-135 phase provides a foundation for unraveling the mechanisms underlying multiple electronic orders in these related materials. Hope this review has captured the rapid advancement in AV$_3$Sb$_5$, offering insights into their complex interplay of charge ordering, superconductivity and other symmetry-breaking phenomena.

~\\
\small
\noindent\textit{Acknowledgments.}
This research is supported by the National Key Research and Development Program of China under Grant No.2024YFA1408100 and National Natural Science Foundation of China under Grant No. 12474131. This research is supported by ``Pioneer'' and ``Leading Goose'' R$\&$D Program of Zhejiang under Grant 2024SDXHDX0007 and Zhejiang Provincial Natural Science Foundation of China for Distinguished Young Scholars under Grant No. LR23A040001. 
This research is supported by the Research Center for Industries of the Future (RCIF) at Westlake University under Award No. WU2023C009.

%\section{Results and discussion}
%\noindent\textbf{Results}

%~\\
%\noindent\textbf{Online content}

%\noindent Supplementary information are available at the online version of the paper. 

~\\
%\noindent \textbf{References}
%\bibliographystyle{naturemag}
\bibliographystyle{CPL}
\bibliography{ReviewKagomeSC}

%\clearpage
%\noindent\textbf{Methods}

%~\\
%\noindent
%\textbf{Growth of single crystals}

%~\\
%\noindent\textbf{Data availability}

%\noindent Data are available from the corresponding author upon reasonable request.

%~\\
%\noindent\textbf{Author contributions}
%T.L. fabricated the devices and did the transport measurements assisted by Z.X., J.W. and Z.L.. J.L. prepared the samples supervised by Z.W. and Y.Y.. Z.P. did theoretical calculations supervised by C.W.. T.L. and X.L. prepared the figures. C.W. and X.L. wrote the paper. X.L. led the project.  All authors contributed to the discussion.

%~\\
%\noindent\textbf{Competing interests}
%The authors declare no competing interests.

%~\\
%\noindent\textbf{Supporting information}
%\noindent\textbf{Additional information}

%\noindent\textbf{Supplementary information} are available at the online version of the paper. 

%\noindent Correspondence and requests for materials should be addressed....

%\clearpage

\end{document}